\documentstyle[psfig]{l-aa}
\def\ros{{\sl ROSAT }}
\def\asca{{\sl ASCA }}

\def\farcs{\hbox{$.\!\!^{\prime\prime}$}}  
\def\approxlt{\mathrel{\hbox{\rlap{\lower.55ex \hbox {$\sim$}}
        \kern-.3em \raise.4ex \hbox{$<$}}}}
\def\approxgt{\mathrel{\hbox{\rlap{\lower.55ex \hbox {$\sim$}}
        \kern-.3em \raise.4ex \hbox{$>$}}}}

\begin{document}
\thesaurus{03         
              (11.01.2;  
               11.09.1;  
               11.09.2;  
               11.19.3;  
               13.25.2)  
}
  \title{\ros HRI discovery of luminous extended X-ray 
    emission in NGC\,6240} 
   \author {Stefanie Komossa \inst{1}, Hartmut Schulz  \inst{2},
\and Jochen Greiner\inst{3}   
  }
\offprints{Stefanie Komossa, \\
skomossa@xray.mpe.mpg.de}
\institute{
Max-Planck-Institut f\"ur extraterrestrische Physik, Postfach 1603,
D-85740 Garching, Germany
\and
Astronomisches Institut der Ruhr-Universit\"at, D-44780 Bochum, Germany
\and
Astrophysikalisches Institut, D-14482 Potsdam, Germany} 
\date{Received: December 1997; accepted: 23 December 1997}
   \maketitle
\markboth{Komossa et al.: Luminous extended X-ray emission in NGC\,6240}
{Komossa et al.: Luminous extended X-ray emission in NGC\,6240}
\begin{abstract}

We report the detection of luminous extended X-ray emission in NGC 6240
on the basis of \ros HRI observations of this ultraluminous 
IR galaxy.  The spatial structure and temporal behavior of the
X-ray source were analyzed.
We find that $\ga 70\%$ of the soft X-ray emission is extended beyond
a radius of 5\arcsec. 
Strong emission can be traced out to a radius of 20\arcsec~
and weaker emission extends out to $\sim$50\arcsec.  
With a luminosity of at least $L_{\rm x} \simeq$
10$^{42}$ erg/s this makes NGC 6240 one of the
most luminous X-ray emitters in {\em extended} emission known.
Evidence for a nuclear compact variable component
is indicated by a 
drop of 32\% in the HRI count rate 
as compared to the PSPC data taken one year earlier.  
No short-timescale variability is detected. 
The HRI data, which represent the first high-resolution
study of the X-ray emission from NGC 6240,
complement previous spectral fits to \ros PSPC data
that suggested
a two-component model consisting of thermal emission
from shocked gas immersed in a starburst wind plus
a powerlaw source attributed to scattered light from an obscured AGN.

We discuss several models to account for the extended and
compact emission.
Although pushed to its limits the starburst outflow model is tenable
for the essential part of the {\em extended} emission.
For the AGN-type component we propose a model 
consisting of a near-nuclear `warm scatterer' that explains
the apparent fading of the X-ray flux within a year as well as
the strong FeK$\alpha$ complex seen in an \asca spectrum.
\end{abstract} 
\section{Introduction}
With a far-infrared 
luminosity of $\sim 10^{12} L_\odot$ (Wright et al.\ 1984) and
a redshift of $z=0.024$,
NGC 6240 is one of the nearest members of the 
class of ultraluminous infrared galaxies
(hereafter ULIRG). The basic, as yet unsolved, enigma
of these objects is the nature of the primary power source
that has to be situated inside the central few arcseconds 
(Wynn-Williams \& Becklin 1993). An amount of
$\sim10^{10} M_{\sun}$ of cold molecular gas
(e.g. Solomon et al. 1997), a
record 2.121$\mu$m-line luminosity from shocked 
`warm' H$_2$ (e.g. van der Werf et al. 1993),
earthbound IR spectra (e.g. Joseph \& Wright 1985, Rieke et al. 1985,
Schmitt et al. 1996), MAMA (Smith et al. 1992) and HST (Barbieri et al. 1993)
observations, 
and recent ISO-SWS spectra (Lutz et al. 1996) all point
towards the presence of hidden prodigious star formation after onset
of a galactic collision, which could
be responsible for the FIR power. Heckman et al. (1987, 1990)
found indications for superwind and supershell activity,
a well-known result of strong starbursts. 

However, the smallness of the recombination line flux
(de Poy et al. 1986), the detection of a  
high-excitation component in HST images
(Barbieri et al. 1995, Rafanelli et al. 1997) and general arguments valid for
ULIRGs as a class (e.g. Sanders et al. 1988) suggest
that a dust-shrouded AGN contributes significantly
to the heating of the dust that emits the FIR radiation.

The unambiguous detection and investigation of an
AGN in NGC 6240 and other interacting ULIRGs would be of prime importance
for our understanding of the formation and evolution
of AGN in general. It has been proposed that starbursts are
the germ cell for the formation of AGN 
(Weedman 1983, Barnes \& Hernquist 1991, Mihos \& Hernquist 1996) 
and interaction may provide the triggers and fuel for both kinds
of activity (see Sanders \& Mirabel 1996 for a recent review).
A large fraction of ULIRGs indeed turned out to be interacting 
systems (e.g. Andreasian \& Alloin 1994, Clements et al. 1996). 

As outlined above, there are indications for a starburst in NGC 6240, 
but what would
be the best evidence for a hidden AGN?
The far-infrared emission is reprocessed black-body like radiation 
containing no direct clue on the nature of the primary source.
Near-IR and mid-IR line spectra provided signatures for a
red giant population and a younger burst. A few high-excitation 
features in IR spectra and in optical HST narrow-band images could be due 
to an
AGN but not necessarily. The optical emission-line spectrum 
(Fosbury \& Wall 1979, Zasov \& Karachentsev 1979, Fried \& Schulz 1983,
Morris \& Ward 1988,
Keel 1990, Heckman et al. 1987, Veilleux et al. 1995,
Schmitt et al. 1996) is 
dominated by LINER-like line ratios over the central $\sim10$ kpc. 
Its large extent and little variation in excitation tracers 
is more easily attributed to
shock-heating rather than to a central photoionizing AGN continuum.

X-rays are an important tool for studying both, an AGN as well as 
starburst components.  
In the \ros band, AGN tend to be dominated by strong powerlaw 
(hereafter PL) emission while starbursts
can usually be represented by thermal spectra.
In a recent analysis
of \ros PSPC spectra from NGC 6240, Schulz et al. (1998; hereafter 
paper I)
found good fits by either a single thermal Raymond-Smith (hereafter RS) 
spectrum
with $L_{\rm 0.1-2.4 keV} = 3.8\,10^{43}$ erg/s (for a distance of 
144 Mpc) 
or a hybrid model consisting of 80\% PL plus 20\% thermal RS 
(dubbed 0.8PL+0.2RS below) contributions and a total
luminosity of $5.2\,10^{42}$ erg/s. Since the spectral shape
with PSPC resolution is not sufficiently distinctive the luminosity 
information
was taken as an additional constraint. 
Due to the 
unprecedented high
luminosity of the single RS model and additional severe
difficulties to explain it in terms of scalable
superbubble models, the hybrid model
was favored. 
This is also supported by the {\asca} detection of a strong FeK$\alpha$ 
line in the X-ray spectrum of NGC 6240 (Mitsuda 1995; the same data indicate
further emission lines around 1--2 keV).  
The powerlaw was attributed to the electron scattered X-ray 
flux from a hidden AGN so that an AGN-plus-starburst scenario 
was proposed for the ultimate power source of NGC 6240 (paper I). 

The deep HRI observations which are discussed below represent the first
high spatial resolution study of the X-ray emission from NGC 6240. 
They allow to 
trace the emission from a thermal starburst source
that is expected to be appreciably spatially extended while an AGN-induced powerlaw
source should be much more compact unless there is extensive large-scale
scattering. Further, they provide information on the long- and short-term
X-ray variability of the source.  
 
Luminosities given below are calculated assuming a distance $d = 144$ Mpc
of NGC 6240. This yields a scale perpendicular to the line of sight
in which
1\arcsec~ 
corresponds to 700 pc in the galaxy.  

\section{Observations and data reduction}
The data analyzed here were taken with the HRI
(Pfeffermann et al. 1987, Zombeck et al. 1990)  
on board of the X-ray satellite \ros 
(Tr\"umper 1983, 
Tr\"umper et al. 1991) and were retrieved from the archive.  
The HRI detector obtains images in the soft X-ray band
(0.1--2.4 keV) at a spatial resolution of about $5\arcsec$. 

Deep HRI X-ray images of NGC 6240 were taken on 
Feb. 24 - March 4, 1994, Aug. 23 - Sept. 15, 1994 and 
Aug. 23 - 25, 1995  
with effective exposure times 
of 5.7, 28.3, and 15.7 ksec, respectively. 

A source-detection procedure was carried out 
with the EXSAS X-ray analysis software 
package (Zimmermann et al. 1994).
In total, 10 X-ray sources 
($>3\sigma$) were detected in 
the field of view.     
The background was determined in a source-free ring around the 
target source. The source photons were selected from a circular region
large enough to ensure that all source photons were included, given the extent
of the source (see Sect. 4). 
NGC 6240 is the brightest source in all pointings and we 
find a background-subtracted mean HRI count rate
of 0.0177 cts/s. 

\section{Temporal analysis}
At first, we have checked for variability in the source flux on longer 
terms,
i.e. between the individual HRI pointings and the earlier PSPC 
observations.
For the individual HRI observations, we find mean source count rates of 
0.0175$\pm{0.0018}$, 0.0179$\pm{0.0008}$ and 0.0174$\pm{0.0010}$ cts/s, 
which is consistent with constant source flux between the respective 
epochs.

During the \ros all-sky survey in 1990, NGC 6240 was detected with a 
PSPC count rate
of $0.086\pm0.016$ cts/s (Voges et al. 1996). 
Converting the count rate from the pointed PSPC observations of 
0.064$\pm{0.005}$ cts/s  
(see paper I; the PSPC observations
were performed in Sept. 1992 and Feb. 1993) 
into
an HRI count rate{\footnote{{the PSPC--HRI count rate conversion
was performed within EXSAS by folding the model spectrum with the detector response matrix
and taking into account the effective area, as well as using the conversion program PIMMS;
both yield the same results.}}  
 (under the assumption of constant spectral shape 
between the observations)
yields a value of   
$CR = 0.025\pm0.002$ cts/s, higher than in the 
HRI data.
This indicates a fading by 
32\% within one year
from the pointed PSPC to the HRI observation. 
We checked the sensitivity 
of 
the counts conversion
to the spectral shape by comparing it for several different models
that provided an acceptable PSPC spectral fit (like a weakly absorbed 
black body
or a strongly absorbed single powerlaw to test extremes) and   
found no significant effect. Then, the conversion of the counts was
performed for the favored 0.8PL+0.2RS hybrid model from paper I 
and, afterwards, this procedure was repeated after 
having halved the PL contribution 
which led to a $\sim$30\% reduction  
of the 
converted counts closely mimicking the observed drop in count rate.

Consequently, a fading of the PL component by nearly a factor of 
two is consistent with the smaller HRI count rates measured 
one year after the last PSPC data had been taken. 

Secondly, we have searched for short time-scale variability
within each HRI observation.
The X-ray lightcurve, binned to time intervals
of 400 s or more, is shown in Fig. \ref{light}.
Within the errors, the lightcurve is consistent
with constant source emission.

  \begin{figure}[thbp]
     \vbox{\psfig{figure=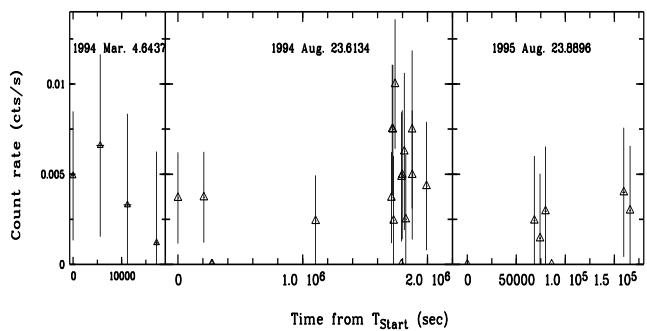,width=8.8cm,height=4.5cm,%
          bbllx=1.4cm,bblly=3.5cm,bburx=12.0cm,bbury=7.6cm,clip=}}\par
\vspace{-0.3cm}
\caption[light]{HRI X-ray lightcurve for the central 5\arcsec~ of 
NGC 6240.
The time is measured in seconds from the start (given in the top of
each panel) of the individual 
observations. The binsize in time was chosen as $\ge$ 400s. 
}
\label{light}
\end{figure}

\section{Spatial analysis}
The X-ray emission from NGC 6240 is found to be significantly extended in all three 
pointed HRI observations. 
For a more detailed analysis of the spatial structure  
we have used the HRI image with the longest exposure time of 28 ksec. 
Numbers given below refer to this observation if not stated otherwise.  
The apparent source extent (not corrected 
for the point spread function) is of the order of at least 25\arcsec~
with evidence for weak emission extending outwards to $\approx$ 
50\arcsec.
Given residual errors in the aspect solution that occasionally produce 
an apparent source extent in HRI-observed sources (David et al. 1992, 
Briel et al. 1994)
which was found to be of the order 1\arcsec~ in FWHM (Morse 1994),   
we note that NGC 6240 extends well beyond 6\arcsec~and 
is the only source in the image for which we find 
significantly extended emission. The other (weaker) sources 
are consistent with a PSF-widened point source. E.g., we find a 
likelihood for source extent
(for its definition see Zimmermann et al. 1994) of $l_{\rm e}$ = 362 for 
the on-axis target source whereas this value is $l_{\rm e}$ = 0
for the second-brightest source in the field and $l_{\rm e}$ = 2.1 for
the source with the second-highest $l_{\rm e}$.  
Finally, we also note that indications for source extent already showed up
in the PSPC observation (paper I, Schulz \& Komossa 1997). 
No point sources embedded in the extended emission were found. 

  \begin{figure}[thbp]
    \vbox{\psfig{figure=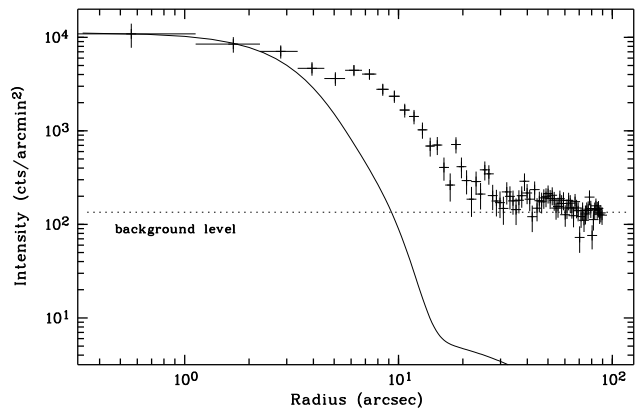,width=8.7cm,
        bbllx=1.7cm,bblly=9.5cm,bburx=11.8cm,bbury=16.5cm,clip=}}\par
\vspace{-0.3cm}
\caption[radial]{Radial profile of the source emission (crosses) as 
compared to the
theoretical PSF for an on-axis point source (solid line). 
Note the logarithmic scale. 
Extended emission is clearly detected. 
}
\label{radial}
\end{figure}
%
The radial profile of the X-ray emission from NGC 6240
is shown in Fig. \ref{radial} together with the instrumental PSF.
Only 31\% of the source photons are found inside a radius of 5\arcsec
(i.e.\ are unresolved by the HRI) whereas 90\% would be expected for a 
point source. This provides an upper limit on the contribution to the X-ray emission  
from a central point source.   

To further analyze the structure of the X-ray image 
we have performed a maximum likelihood (ML; Cruddace et al. 1988)
analysis (Greiner et al. 1991). 
In this approach, a smoothed background image  
is first produced. 
Then, the point spread function of a source with a width appropriate
for the corresponding off-axis angle is fitted to the 
background-subtracted source photons 
using a spatial grid of spacing 2$\farcs$5. 
The likelihood for the existence of a source is calculated for
each grid point.  
The advantage 
of this method is the strong suppression of features which are smaller 
than
the PSF while structures wider than the PSF are contrast enhanced (due to 
the steep 
likelihood variation as a function of counts) as compared to a simple 
count rate 
image.  
\begin{figure*} 
 \vbox{\psfig{figure=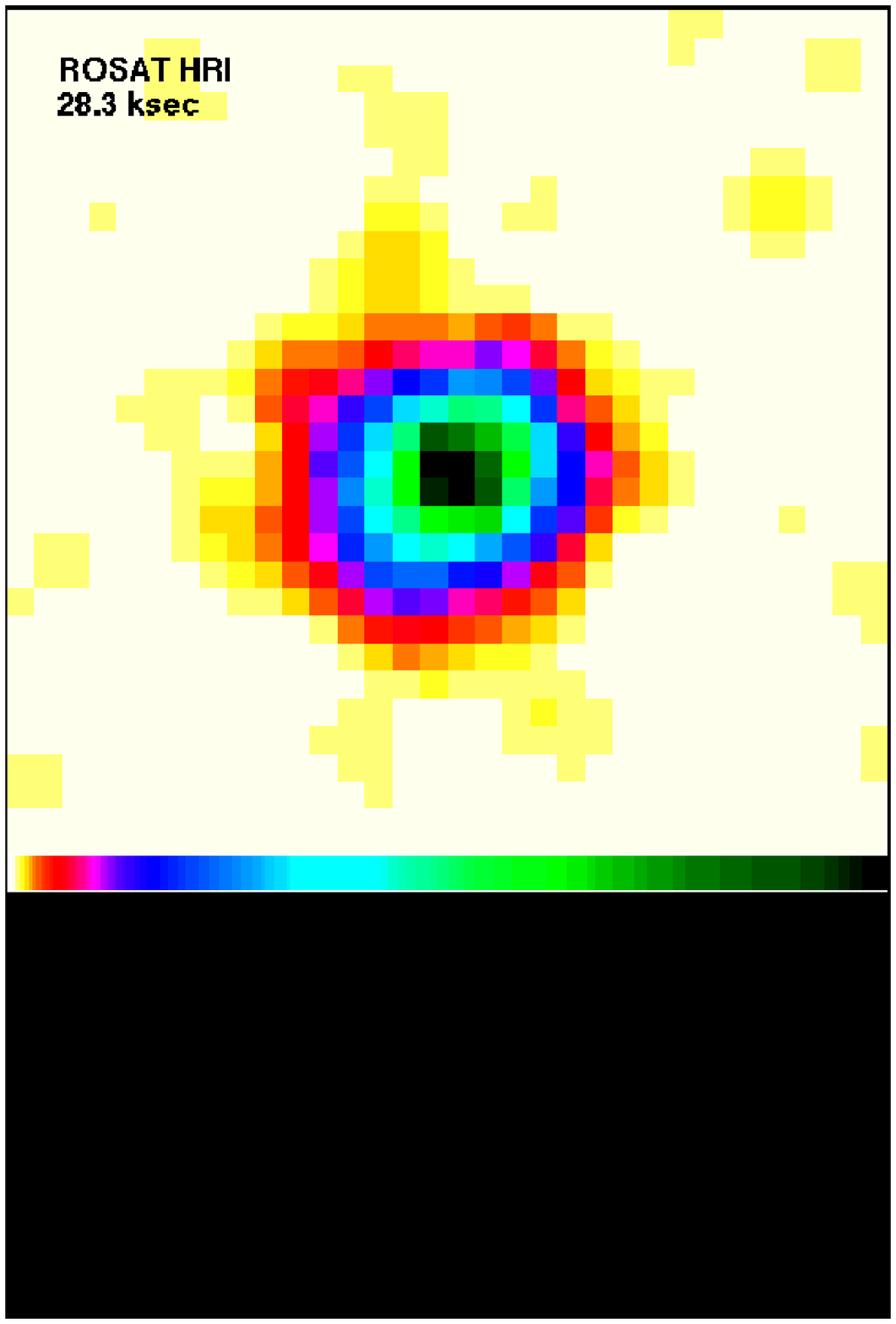,width=8.5cm,
  bbllx=1.5cm,bblly=8.0cm,bburx=15.9cm,bbury=22.5cm,clip=}}\par
\hfill
\begin{minipage}[]{0.48\hsize}\vspace{-3.0cm}
\hfill
\caption[ml]{Maximum likelihood image of the X-ray emission of NGC 6240
produced as described in the text. 1 pixel corresponds to a scale of 
2$\farcs$5.
}
\label{ml}
\end{minipage}
\vspace{-0.3cm}
 \vbox{\psfig{figure=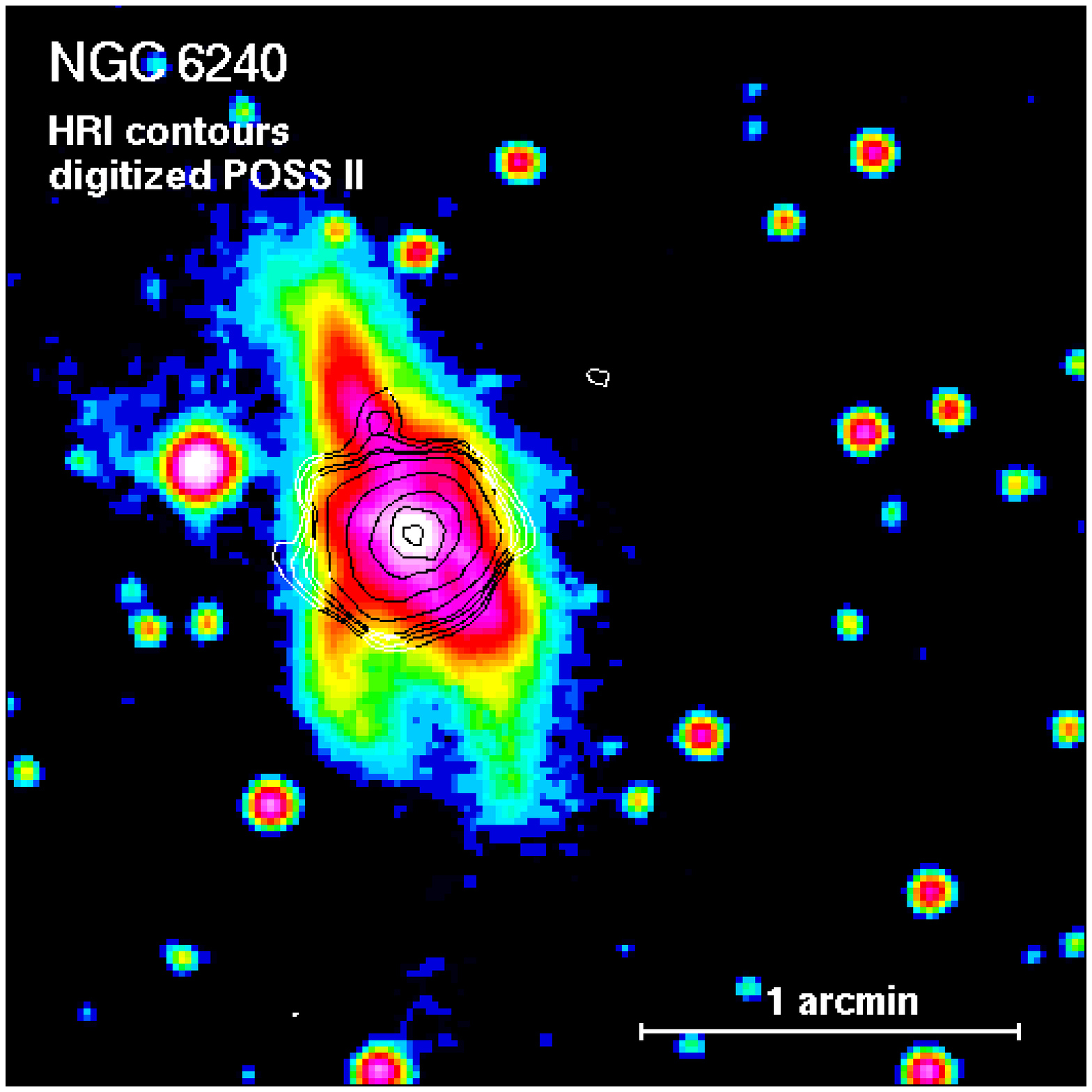,width=13cm,
  bbllx=2.3cm,bblly=10.45cm,bburx=18.1cm,bbury=28.1cm,clip=}}\par
\hfill
\begin{minipage}[]{0.21\hsize}\vspace{-6.0cm}
\hfill
\caption[over]{Overlay of the ML-deconvolved X-ray contours 
on the optical image of NGC 6240 from the digitized Palomar sky survey 
(POSS) II.
The contours are shown for likelihood values of $l$ = 5.8, 
8.0 (3$\sigma$), 12.5, 16.6, 33.2, 83, 207, 348, and 498. 
}

\label{over}
\end{minipage}
\end{figure*}
%
The ML image is shown in Fig. \ref{ml}.
The X-ray contours for this ML image are superimposed on an optical
image of NGC 6240 in Fig. \ref{over}. 
Due to the boresight error of the 
X-ray 
telescope
there is a systematic uncertainty in source position of order 10\arcsec~(Briel 
et al. 1994). 
This causes a displacement of the X-ray emission maximum 
relative to the optical maximum. Within this 10\arcsec~error, the X-ray positions  
are consistent 
with the position of the optical maximum.
The shift was corrected for in Fig. \ref{over}. 

\section{Discussion}
\subsection{Time variability}

The search for variability in the data sets
is of particular importance, since its detection
would either imply a direct view of an active source
or scattering on a small spatial scale.
Constant source flux would be consistent with the superwind 
interpretation or a `hidden' AGN the radiation of
which is scattered on a more extended scale. 

We do not find significant variability on short timescales  
but, as
outlined in Sect.\ 3, there are indications
for a 30\% variation between the PSPC observations finished in Feb.\ 
1993 and 
the HRI observations commencing in Feb.\ 1994. In any simple model, 
this limits the size of an appreciable part of the source plus 
possible 
scattering mirror to less than one light year.
In paper I, in which no variability among the PSPC observations was found,
we estimated the effect of scatterers of sizes ranging from about 600 to 
30\,000 lys.
If the variation is real such large-scale scatterers cannot play
the dominant role. Therefore,  below (Sect. 5.4) we investigate in more
detail the alternative possibility of a compact scatterer.

\subsection{Source extent and structure}

The exciting new discovery of the HRI observations is that the soft X-ray
source in NGC 6240 is appreciably extended.
For a point source, 90\% of the photons are expected within the
central radius of 5\arcsec~ (the resolution limit), whereas we find only
30\%. At face value, this limits the contribution of a central core source
to 33\%.

This percentage estimate only applies to the count rate, though, not
to the flux contribution of a compact component.
For instance, in our finally
discussed hybrid model of paper I the (presumably compact) powerlaw
component contributes 80\% of the total {\em flux}, but only 
60\% in {\em count rate}.
Major reasons for the lower percentage in count rates are the
different spectral shapes and the 
relatively stronger effect of the absorption by cold gas on the powerlaw
than on the thermal Raymond-Smith spectrum.
When attributing the above 
discussed drop of the total count rate by 32\% to the powerlaw component 
the flux
had to decrease to 41\% which would lead
to a 34\% contribution of
the HRI counts inside 5\arcsec. Since 30\% are observed  
the 0.8PL+0.2RS model fitted to the earlier-epoch PSPC data is consistent
within the errors with the new HRI data. 

The X-ray emission can be traced out to a distance of at least 20\arcsec.
Scenarios for the origin of this emission are discussed in the next section.  
On a weak emission level, several `fingers' are apparent that extend
outwards and could be related to the tidal
interaction. 

\subsection{Interpretation of the extended structure}

The basic challenge for any model of the extended component is the huge
luminosity required, at least 10$^{42}$ erg/s in the soft X-ray band. 
This limit derives from PSPC spectral fits performed in paper I.
None gave $L_{\rm x}$ below
$2\,10^{42}$ erg/s. In order
to determine the minimal luminosity consistent with the data
we carried out further spectral fits.
E.g., we fixed the cold column to the Galactic value
and ran a two-component RS fit leading to a total luminosity of
$3\,10^{42}$ erg/s.
Even when taking into account the 32\% lower count rate in the HRI
observation,
we consider $1\,10^{42}$ erg/s as a well established lower limit
of the luminosity.

Since NGC 6240 is a merger, presumably the intermediate product
of the
collision of two gas rich disk galaxies, one might ask whether the
shock converted kinetic energy of the interstellar media of both
galaxies
could have heated the X-ray nebula. Converting $Mv^2/2=10^{58}$ erg
(with $M = 10^{10} M_{\sun}$ and $v=300$ km/s) completely into
the X-ray luminosity would suffice for $10^{7-8}$ years. This number fits
because it is about the dynamical time scale of the collision.
However, this energy could not sustain the total luminosity of
the galaxy of $10^{12} L_{\sun}$ within
the time it is generated (only for $10^5$ years).
Hence, the power source cannot be drawn from the collision.

In paper I  
we already noted that  
electron scattering of a nuclear component, e.g.
from an obscured AGN, did not appear feasible on a 10 kpc scale.
Scattering
on small scales will be discussed in the next section. With large-scale
inverse Compton scattering  
an {\rm extended} PL component could
be produced in principle but the relevant parameters
(magnetic fields and diffusion lengths)
estimated from the measured radio sources did not provide positive
support for such a scenario (Colbert et al. 1994).

When relating the X-ray nebula to the nuclear energy source
the most obvious candidate is the hidden starburst discussed in the
introduction. A starburst evolves by first forming a cavity in the ISM of
the host galaxy created by the accumulating winds of the early-type
stars. Subsequently the hot gas develops a kpc-sized
expanding superbubble that is fed by 
supernova explosions
generated after the first few million years. The bubble is filled with
shock-heated thin gas surrounded by a cold shock-compressed `supershell'
formed from
swept-up ISM. The shell will be stable as long as the bubble expands
decelerating. Evaporation will take place at the interface of the cold
shell and
hot cavity leading to a radial temperature and density distribution
described analytically by Mac Low \& McCray (1988; for numerical 
hydrodynamical simulations with specific initial conditions
see e.g. Tomisaka \& Bregman 1993, Suchkov et al. 1994).
The theory is parameterized by the mechanical input power (given by the
supernova rate), the initial ISM density and the expansion time. The
corresponding 
equations given in paper I 
showed that
from a minimum of shocked H$\alpha$ emission
or from the derived supernova rate one obtains a mechanical input power of
$L_{\rm mech}=3\,10^{43}$ erg/s which, within $3\,10^7$ yrs, can
drive a single
shell to an extent $R \sim 10$ kpc within a medium of $0.1$ cm$^{-3}$
particle density.

While the bubble size was guessed from uncertain H$\alpha$ imaging in
paper I
it is now possible to derive the X-ray size from the HRI data.
Figs. \ref{radial},\ref{over} show that $R = 20 \arcsec$
corresponding to 14 kpc encloses the X-ray emission on the $3\sigma$ level.
It is easy to increase the originally assumed 10 kpc - extent
by 25\% via
increasing $L_{\rm mech}$  by a factor of 3 (still compatible with
NGC 6240) or by 50\% via doubling the time scale
without getting into conflict with other observed parameters. 

There is  still some
faint emission  beyond 14 kpc
below the $3\sigma$-level. Only speculations about its origin
are possible with present data. In vein of the supershell scenario,
it could be old halo emission
from previous bursts.

With a soft X-ray luminosity of at least $L_{\rm x} \simeq 10^{42}$ erg/s 
the {\em extended} emission in NGC 6240 is one of the most
luminous extended X-ray emitters known. 
For comparison, starburst
galaxies typically show
$L_{\rm x} \simeq 10^{39}$ erg/s with a range between $10^{38}$ - $10^{41}$ erg/s
(e.g. Fabbiano 1989, Vogler 1997); the total soft X-ray emission of the ultraluminous
IR galaxy Arp\,220 is of order $4\,10^{40}$ to $2\,10^{41}$ erg/s
(depending on the value of cold absorption; Heckman et al. 1996)
and the extended emission in the Seyfert\,2 galaxy NGC 4388 is about
$3\,10^{40}$ erg/s (Matt et al. 1994).
It is interesting to note that despite early reports for a class 
of very X-ray luminous, but apparently inactive, galaxies, optical 
follow-up observations 
revealed AGN tracers in all objects examined (Moran et al.\ 1994,
1996, Wisotzki \& Bade 1997) which led to the suggestion that X-ray
emission above 10$^{42}$ erg/s should always be attributed to an AGN.
The extended emission in NGC 6240 slightly exceeds this limit. 

Could the {\em whole} X-ray emission of NGC 6240 be due to shocked supershells?
We already noted in paper I that lowering
the PL contribution too much tends to strongly increase the
luminosity demand for the extended component.
For instance, a pure Raymond-Smith fit was considered improbable because
in this case
the thermal X-ray nebula would have to emit outstandingly powerful
$10^{10} L_{\sun}$ in the \ros band which cannot be accommodated  with
any scalable superbubble model.

\subsection{Contribution from a very compact component: AGN reflection mirror}
Referring to the {\ros} PSPC and {\asca} evidence for the contribution of 
a powerlaw
component to the X-ray spectrum that represents scattered light from a 
'hidden'
AGN, some estimates for scattering in the ambient ionized gas on scales 
larger than a few hundred lyrs were presented in paper I. 
Such extended scatterers would, however,
be unable to allow for a notable flux variation within such a short
timescale like a year.
This could only be achieved by a directly seen AGN or a
sufficiently small scatterer very close to the continuum source to
ensure a high covering factor or a collection of scatterers at larger 
distances
from the continuum source but arranged in a way
that light travel times to the observer are in close agreement. The last
solution appears contrived.
 
Lacking any straightforward evidence for a significant optical
nonstellar continuum an unextinguished direct view of the
nonthermal continuum source appears precluded. 
An {\asca} spectrum
reveals an FeK$\alpha$ line complex with an
equivalent width of $\sim 2$ keV which indicates the presence of an AGN 
(Mitsuda 1995). 
Such a line  is not generally observed in starbursters (e.g. Ptak et al. 1997). 
However, in Seyfert 1 galaxies FeK$\alpha$ is
usually believed to arise via Compton reflection from an
accretion disk yielding an equivalent width
of only a few hundred eV with respect to the original continuum and could
only reach a few keV with respect to the reflected continuum.
If one looks face-on towards the accretion disk the original
nonthermal continuum should be visible as well.  Consequently, we have
to locate the scatterer elsewhere. 

The requirements for a compact scatterer are a size $R \approxlt 1$ ly 
(from the temporal variability) and that 
for an efficient effect (to lower geometrical demands) the
electron scattering optical depth $R n_e \sigma_e$ ($\sigma_e = 
6.65\,10^{-25}$ cm$^2$)
should not be too much below unity. These conditions lead to
$n_e \approxgt 1.5\,10^6$ cm$^{-3}$. 
This component may be identified with a molecular torus and/or 
warm gas in the vicinity of the nucleus (so-called `warm absorber';
e.g. Pan et al. 1990, Otani et al. 1996, Komossa \& Fink 1997). 
This material will appear as a `warm scatterer' under appropriate geometrical
conditions.

We have tested this possibility by computing a simple model for such a
highly ionized scatterer and emitter  
using the photoionization code {\em Cloudy} (Ferland 1993). The input continuum is
our standard AGN continuum (e.g. Komossa \& Schulz 1997) with an X-ray powerlaw
of photon index $\Gamma = -1.9$ (defined as in 
$\Phi_{\rm photon} \propto E^{\Gamma}$). 
Solar abundances (Grevesse \& Anders 1989)
were assumed. 
With a column density of the reflector of $N_{\rm w} \simeq 23.0$ and 
an ionization parameter of $\log U \simeq 1.0$\footnote{these values were 
found
for an explicit single-component fit of the model of a `warm reflector' to the 
\ros PSPC
spectrum (Komossa 1997)  
neglecting the contribution of the 
extended component; we  only intend
to show trends here}}   
we find strong intrinsic emission in FeK$\alpha$ due to recombination
processes with a contribution
from several 
ionization stages.  
We predict an 
equivalent width of $\sim$6 keV relative to the reflected continuum. The diffuse 
continuum
at these photon energies from the warm gas is negligible compared to the 
reflected
continuum. Hence, it is possible to fit the \ros spectrum with such a 
warm reflector
and produce FeK$\alpha$ with an equivalent width of several keV.
We also predict the presence of a (weak) Fe edge in the reflected
spectrum and further line emission around 1--2 keV. 
For a more detailed study, in which also the
abundance of the material could be varied, 
high spectral resolution X-ray observations (e.g. with AXAF) 
are needed. 

  \begin{figure}[thbp]
      \vbox{\psfig{figure=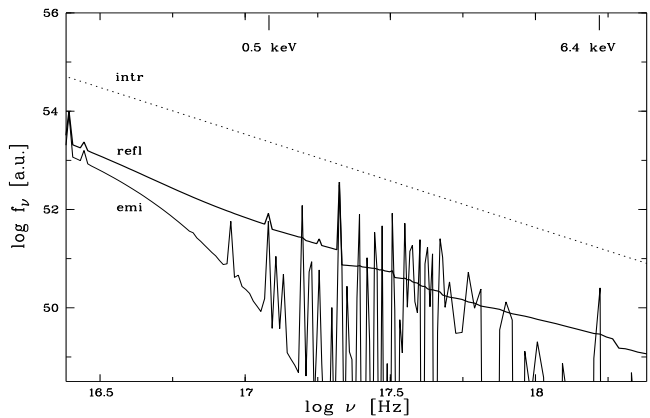,width=8.7cm,height=5.5cm,%
          bbllx=2.9cm,bblly=1.1cm,bburx=18.1cm,bbury=12.2cm,clip=}}\par
      \vspace{-0.3cm}
 \caption[feka]{Spectral components of the warm reflector. The incident
continuum is shown as dotted line. The thin solid line corresponds to 
the emitted spectrum and the thick solid line to the reflected spectrum.   
The abscissa brackets the energy range 0.1 -- 10 keV.  
}
\label{feka}
\end{figure}

\subsection{What powers the FIR emission of NGC 6240 ?} 

Given the strong evidence for both, a super-starburst as well as an
AGN in NGC 6240, we may ask which component provides the ultimate
power source for  
the FIR emission of this ULIRG{\footnote{Lutz et al. (1996)
addressed this question on the basis of ISO data, and conclude 
that the strengths of low-ionization emission lines can be consistently
explained by {\em starburst} models.}}.

From the mechanical power used above, models given in
Leitherer \& Heckman (1995) predict $L_{\rm bol}$(Starburst) to be all or
a major fraction of the $10^{12} L_{\sun}$ which NGC 6240 emits. Several
$10^{10} M_{\sun}$ of dense molecular gas are available in the central kpc
of NGC 6240 to fuel the super-starburst.
However, although the consistency is gratifying the theory applied
is much too idealized and too boldly upscaled to be watertight.
 
In fact, the AGN component may provide enough power as well:
Given an X-ray luminosity in scattered emission of a few 10$^{42}$ erg/s
one obtains an {\em intrinsic} luminosity of order 10$^{44-45}$ erg/s, 
depending on the covering factor of the scattering material. 
This is, again, an appreciable fraction of the FIR luminosity.  
    
So it seems that in case of NGC 6240, both components contribute
in comparable strength. 

\section{Summary and Conclusions}
We detected luminous extended X-ray emission in NGC 6240 in \ros HRI 
data. 
At the given spatial resolution
the source looks nearly spherical 
and contains its 
most
significant emission 
within a radius of
20\arcsec~(or 14 kpc for a distance d=144 Mpc) where the total
0.1--2.4 keV X-ray luminosity amounts to at least $\sim 10^{42}$ erg 
s$^{-1}$.
At the epochs of the observations 
at most 40\% of this luminosity arises within the innermost
region of 5\arcsec~radius. 

The extended emission can be consistently described by crude supershell 
models
thereby explaining it as the result of a super-starburst with a total
luminosity close to $10^{12} L_{\sun}$.

The presence of an additional compact AGN component is in X-rays indicated
by (i) a decrease in the count rate between Feb. 1993 (last PSPC 
observation)
and Feb. 1994 (first HRI observation) by 32\%, (ii) a probable
powerlaw component necessary to fit PSPC spectra and (iii) a strong
FeK$\alpha$ complex detected in \asca spectra. We propose
a model in which near-nuclear  
warm 
gas ionized by the AGN powerlaw continuum emits FeK$\alpha$ which
is seen superposed on the reflected continuum.  

Both components, the starburst as well as the AGN provide enough
power to explain the luminous FIR emission in NGC 6240 and it seems
that both contribute with comparable strength.   
\begin{acknowledgements}
We thank Joachim Tr\"umper for a critical reading of the manuscript,
Gary Ferland for providing {\em Cloudy}, and Andreas Vogler for help in 
plotting the overlay contours.  
The \ros project is supported by the German Bundes\-mini\-ste\-rium
f\"ur Bildung, Wissenschaft, Forschung und Technologie (BMBF/DLR) and
the Max-Planck-Society.
JG is supported by  
BMBF/DLR under contract No.
FKZ 50 QQ 9602 3.
The optical image shown is based on photographic data of the 
National Geographic Society -- Palomar
Observatory Sky Survey (NGS-POSS).
\end{acknowledgements}

\end{document}